\documentclass{elsart}
\usepackage{natbib}
\usepackage{graphicx}

\usepackage{amssymb}
\usepackage{amsmath}
\usepackage{bm}

\begin{document}
\newcommand {\ds}{\displaystyle}
\def\sun{\hbox{$\odot$}}
\def\hinvmpc{h^{-1}\,{\rm Mpc}}
\def\hmpcinv{h\,{\rm Mpc}^{-1}}
\newcommand{\simgt}{\lower.5ex\hbox{$\; \buildrel > \over \sim \;$}}
\newcommand{\simlt}{\lower.5ex\hbox{$\; \buildrel < \over \sim \;$}}

\begin{frontmatter}
\title{Calibrating the Nonlinear Matter Power Spectrum: Requirements 
for Future Weak Lensing Surveys}

\author[address1]{Dragan Huterer},
\address[address1]{Kavli Institute for Cosmological Physics and the
Department of Astronomy and Astrophysics, University of Chicago, 
Chicago, IL~~60637}
\author[address2]{Masahiro Takada},
\address[address2]{Astronomical Institute, Tohoku University, Sendai
 980-8578, Japan}
%\thanks[dhemail]{E-mail: dhuterer@kicp.uchicago.edu}
%\thanks[mtemail]{E-mail: takada@astr.tohoku.ac.jp}

\begin{abstract}
Uncertainties in predicting the nonlinear clustering of matter are among the
most serious theoretical systematics facing the upcoming wide-field weak
gravitational lensing surveys.  We estimate the accuracy with which the matter
power spectrum will need to be calibrated in order not to contribute
appreciably to the error budget for future weak lensing surveys. We consider
the random statistical errors and the systematic biases in $P(k)$, as well as
some estimates based on current N-body simulations.  While the power spectrum
on relevant scales ($0.1\lesssim k/\hmpcinv \lesssim 10$) is currently
calibrated with N-body simulations to about 5-10\%, in the future it will have
to be calibrated to about 1-2\% accuracy, depending on the specifications of
the survey.  Encouragingly, we find that even the worst-case error that mimics
the effect of cosmological parameters needs to be calibrated to no better than
about 0.5-1\%. These goals require a suite of high resolution N-body
simulations on a relatively fine grid in cosmological parameter space, and
should be achievable in the near future.
\end{abstract}
%\maketitle
\end{frontmatter}

\section{Introduction}

Weak gravitational lensing (WL) of distant galaxies is determined by the
total matter overdensity along the line of sight. In more technical terms,
weak lensing measures the matter power spectrum weighted with a
function that depends on the geometry of the lens-deflector-observer
system. Therefore, in order to predict the WL signal for a given set
of cosmological parameters, one needs to be able to accurately
calibrate the matter power spectrum $P(k)$.

The sensitivity of weak lensing surveys peaks at scales of about 10 arcminutes on
the sky, corresponding to intervening structures of size $\sim 1 \hinvmpc$ at
redshift $\sim 0.5$, which is approximately where the lensing efficiency of
distant galaxies peaks (e.g.\ Jain \& Seljak 1997).  Therefore, the scales
probed are {\it nonlinear} and need to be calibrated by N-body simulations.
Unfortunately, one cannot afford to restrict the measurements to linear scales
for which there are accurate analytical predictions (and which corresponds to
scales $\gtrsim 1$ deg.\ at $z\sim 0.5$), since the information from large
scales only is poorly constraining due to cosmic variance (e.g.\ Huterer 2002).

Therefore, accurate calibration of the fully nonlinear matter power spectrum
$P(k)$ is necessary to ensure that the cosmological parameter determination
from weak lensing surveys is not compromised.  Current accuracy in the matter
power spectrum at mildly nonlinear scales is about 5-10\% (White \& Vale 2004;
Heitmann et al. 2004) which is sufficiently small for current surveys (for
reviews of weak lensing, see e.g., Bartelmann \& Schneider 2001, Van Waerbeke
\& Mellier 2003, Refregier 2003). However, the accuracy will need to be better
for future wide-field surveys.  The accuracy and agreement of various N-body
simulations is rapidly improving and the currently popular fitting formulae for
the power spectrum (e.g.\ Ma et al.\ 1999, Jenkins et al.\ 2001, Smith et
al. 2003) will soon be insufficiently accurate and flexible and one will need
to resort to a full suite of N-body simulations, interpolating on a grid of
cosmological models (and thereafter, perhaps, building a new set of more
flexible and accurate fitting formulae).

In this work we consider the requirements of future deep, wide-field WL surveys
such as Dark Energy Survey\footnote{http://cosmology.astro.uiuc.edu/DES},
PanSTARRS\footnote{http://pan-starrs.ifa.hawaii.edu}, Supernova/Acceleration
Probe (SNAP\footnote{http://snap.lbl.gov}; Aldering et al.\ 2004) and Large
Synoptic Survey Telescope (LSST\footnote{http://www.lsst.org}). For the
fiducial model we assume sky coverage of 1000 sq. deg. with 100 galaxies per
arcmin$^2$. We vary these numbers later to determine the robustness of our
results to survey specifications. We assume a flat universe with matter energy
density relative to critical $\Omega_M=0.3$, dark energy equation of state
$w=-1$, and power spectrum normalization $\sigma_8=0.9$.  We use the spectral
index and physical matter and baryon energy densities with mean values $n=1.0$,
$\Omega_M h^2=0.140$ and $\Omega_B h^2=0.023$ respectively.  The parameters
whose future accuracy can most benefit from future WL surveys are $\Omega_M$,
$w$ and $\sigma_8$, and we pay particular attention to these three.  We do not
add any priors whatsoever, since we are considering the efficacy of WL surveys
in their own right; moreover, we have checked that reasonable priors (for
example, those based on expected accuracies from the Planck experiment) have
minimal effect on the three aforementioned parameters and all of our
results. Note that we do not consider the time variation in $w$ since the
best-measured mode of an arbitrary $w(z)$ is about as well measured as $w={\rm
constant}$ (Huterer \& Starkman 2003), and it is this particular mode, being
the most sensitive to theory systematics, that will drive the accuracy
requirements.  Hence, it is sufficient to consider the constant $w$ case.
Throughout we consider lensing tomography with 10 redshift bins equally spaced
between $z=0$ and $z=3$ and use the lensing power spectra on scales $50\le \ell
\le 3000$. Over this range of multipoles, statistical properties of the
lensing fields are nearly Gaussian and, furthermore, complex baryonic
effects that are expected to be significant on smaller scales are strongly suppressed
 (e.g.\ White 2004, Zhan \& Knox 2004).  [Cosmological parameter extraction
from WL tomography has been extensively discussed in the past, e.g.\ Hu 1999, Huterer 2002,
Hu 2003, Refregier et al. 2003, Takada \& Jain 2004, Takada \& White 2004, Song
\& Knox 2004, Ishak et al. 2004.]

The convergence power spectrum at a fixed multipole $\ell$ 
and for the $i$th and $j$th tomographic bin  is given by
\begin{equation}
P_{ij}^{\kappa}(\ell) = 
%{\ell^3\over 2\pi^2} 
\int_0^{\infty} dz \,{W_i(z)\,W_j(z) \over r(z)^2\,H(z)}\,
 P\! \left ({\ell\over r(z)}, z\right ),
\label{eq:pk_l}
\end{equation} 
\noindent where $r(z)$ is the comoving angular diameter distance and $H(z)$ is
 the Hubble parameter.  The weights $W_i$ are given by $W_i(\chi) = {3\over
 2}\,\Omega_M\, H_0^2\,g_i(\chi)\, (1+z)$ where $g_i(\chi) =
 r(\chi)\int_{\chi}^{\infty} d\chi_s n_i(\chi_s) r(\chi_s-\chi)/r(\chi_s)$, $\chi$ is
 the comoving radial distance and $n_i$ is the comoving density of galaxies if
 $\chi_s$ falls in the distance range bounded by the $i$th redshift bin and
 zero otherwise.  We employ the redshift distribution of galaxies of the form
 $n(z)\propto z^2\exp(-z/z_0)$ that peaks at $2z_0=1.0$ (for LSST, we use
 $2z_0=0.7$).
The observed convergence power spectrum is
\begin{equation}
C^{\kappa}_{ij}(\ell)=P_{ij}^{\kappa}(\ell) + 
\delta_{ij} {\langle \gamma_{\rm int}^2\rangle \over \bar{n}_i},
\label{eq:C_obs}
\end{equation}
\noindent where $\langle\gamma_{\rm int}^2\rangle^{1/2}$ is the rms intrinsic
shear in each component which we assume to be equal to $0.22$, and $\bar{n}_i$
is the average number of galaxies in the $i$th redshift bin per steradian.  The
cosmological constraints can then be computed from the Fisher matrix
\begin{equation}
F_{ij} = \sum_{\ell} \,{\partial {\bf C}\over \partial p_i}\,
{\bf Cov}^{-1}\,
{\partial {\bf C}\over \partial p_j},\label{eq:latter_F}
\end{equation}
\noindent where ${\bf Cov}^{-1}$ is the inverse of the covariance matrix 
between the observed power spectra whose elements are given by
\begin{equation}
{\rm Cov}\left [C^{\kappa}_{ij}(\ell), C^{\kappa}_{kl}(\ell')\right ] = 
{\delta_{\ell \ell'}\over (2\ell+1)\,f_{\rm sky}\,\Delta \ell}\,
\left [ C^{\kappa}_{ik}(\ell) C^{\kappa}_{jl}(\ell) + 
  C^{\kappa}_{il}(\ell) C^{\kappa}_{jk}(\ell)\right ].
\label{eq:Cov}
\end{equation}
assuming that the convergence field is Gaussian, which is a good
approximation at $\ell<3000$.  We alternatively consider a SNAP-type survey (1000
sq.\ deg., 100 gal/arcmin$^2$) and an LSST-type survey (15000 sq.\ deg., 30
gal/arcmin$^2$). The fiducial marginalized accuracies in $\Omega_M$, $w$
and $\sigma_8$ for the two surveys,  without any theoretical systematics, are
given in Table 1.
%The fiducial SNAP-type survey, without any theoretical systematics,
%determines $\Omega_M$, $w$ and $\sigma_8$ to accuracies of about 0.008, 0.05
%and 0.006 respectively. The LSST-type experiment (15000 sq.\ deg., 30
%gal/arcmin$^2$), under the same assumptions, has accuracies in the three
%parameters that are about a factor of two smaller.

\begin{table}
\begin{center}
\begin{tabular}{||c|c|c||}
\hline\hline
Parameter  
        & \rule[-2mm]{0mm}{5mm} SNAP Error 
        & \rule[-2mm]{0mm}{5mm} LSST Error 
	\\\hline
$\Omega_{\rm DE}$    & 0.008  & 0.003  \\\hline
$w$                  & 0.052  & 0.025  \\\hline
$\sigma_8$           & 0.006  & 0.003  \\\hline
\hline
\end{tabular}
\caption{Fiducial 1-$\sigma$ errors in $\Omega_M$, $w$ and $\sigma_8$ for
the SNAP-type and LSST-type surveys {\it assuming no systematics}.
These errors were obtained by marginalizing over all (five) other  cosmological parameters.
}
\end{center}
\end{table}

In Section \ref{sec:stat} we assume random errors in the matter power spectrum $P(k)$
and determine their maximal allowed size so that they do not appreciably add to
the overall error budget in future surveys. In Section \ref{sec:sys} we explore
biases due to several specific sources of error, and also consider the
worst-case scenario when errors in $P(k)$ conspire to maximally bias the cosmological
parameters. We conclude in Section \ref{sec:conclusions}.

\section{Statistical errors in $P(k)$}\label{sec:stat}

First, let us parameterize the deviations in the matter power spectrum around
its fiducial value by 30 parameters $\delta \ln P_i$ in bins 
spaced equally in $\log k$ from $k=0.1\hmpcinv$ to $k=10\hmpcinv$. More precisely,
if $k$ is in the $i$th bin, we allow the variations
\begin{equation}
P(k, z) \rightarrow P(k, z)\left (1+\delta\ln P_i\right )
\end{equation}
\noindent where $\delta  \ln P_i$ are assumed to be independent gaussian
random variables, not
determined by the cosmological parameters, with mean zero.  Note too that we
assume that relative variations in $P(k, z)$, parametrized by $\delta\ln P_i$, are
independent of $z$. This simplifying assumption is motivated by the fact that
the source of error in $\ln P(k, z)$ is imperfect accuracy of N-body
simulations, where biases in $\ln P(k, z)$ are likely to be weakly dependent on
$z$ in the range relevant for WL.  We compute the linear power spectrum using
the fitting formulae of Eisenstein \& Hu (1999) which were fit to the numerical
data produced by CMBFAST (Seljak \& Zaldarriaga 1996). We generalize the
formulae to $w\neq -1$ by appropriately modifying the growth function of
density perturbations. To complete the calculation of the full nonlinear power
spectrum we use the fitting formulae of Smith et al.\ (2003).

\begin{figure}
\begin{center}
\includegraphics[height=3.0in, width=4.in]{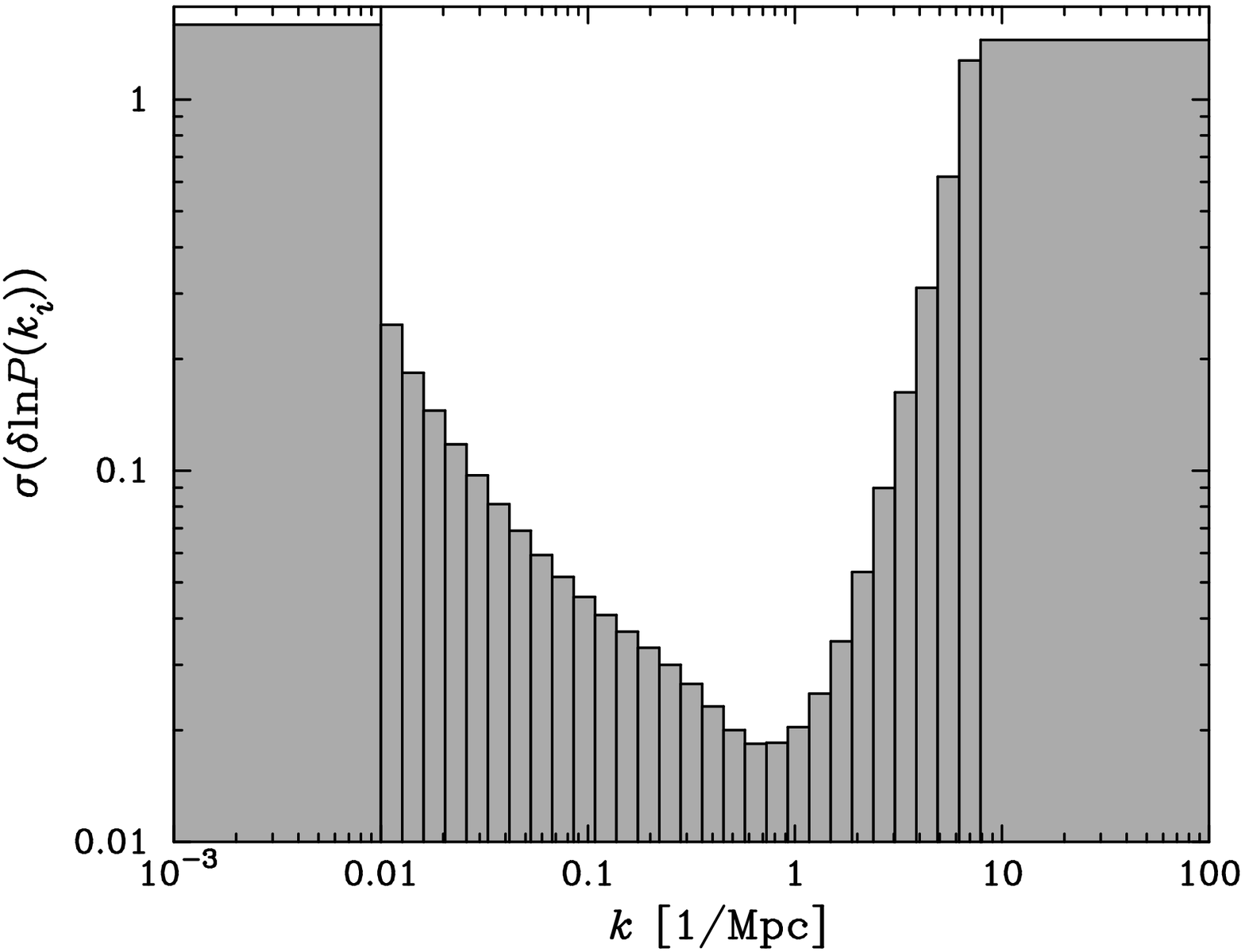}
\includegraphics[height=3.0in, width=4.in]{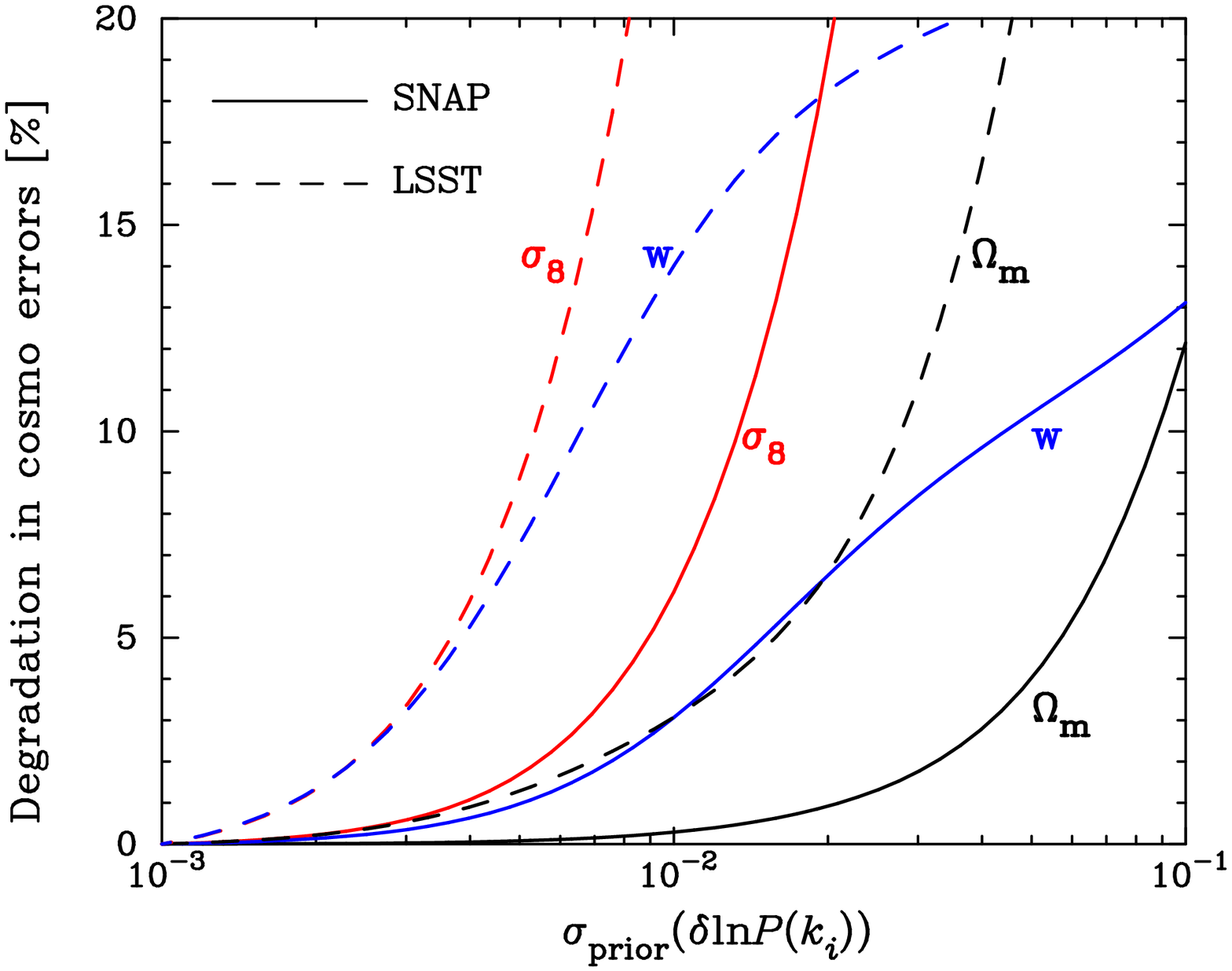}
\end{center}
\caption{Effects of random statistical error on the power spectrum $P(k)$.  In
the top panel we represent $P(k)$ in 30 bins distributed uniformly in $\log k$
from $k=0.1$ to $10\hmpcinv$ and show the unmarginalized accuracy in each band
power,  $\delta \ln P_i$, assuming that the cosmological parameters are perfectly
known. It is clear that the scale that is most sensitive to weak lensing
measurements is $k\sim 1\hmpcinv$, and therefore $P(k)$ around this scale needs
to be calibrated most accurately. We also find that about 10 band powers of
$P(k)$ in the plotted range are independent.  The bottom panel shows the
increase in the error in cosmological parameters $\Omega_m$, $w$ and $\sigma_8$
as a function of prior imposed on each of the 10 nuisance parameters $\delta
\ln P_i$.  The solid and dashed curves are the results for SNAP- and LSST-type
surveys, respectively.  The required accuracy is about $2\%$ from $\sigma_8$
and 5-10\% from $\Omega_M$ and $w$ for SNAP, and about a factor of two more
stringent for LSST.  }
\label{fig:Pk_prior}
\end{figure}

We assume for a moment that all cosmological parameters are precisely known and
fixed.  The top panel of Figure \ref{fig:Pk_prior} shows that the
unmarginalized accuracies in $\delta \ln P_i$ range from a few hundredths
to a few tens, except on very small scales. Not surprisingly, the best
accuracy is achieved on scales of $\sim 1\hmpcinv$ to which the lensing power
spectra of our interest are most sensitive.  From the covariances between
different $\delta \ln P_i$ (i.e.\ the off-diagonal components of the Fisher
information matrix), we find that only about 10 bins in $P(k)$ are independent
-- in other words, the weak lensing kernel is broad enough to differentiate
only around 10 parameters in $P(k)$ in this range of $k$.  We then investigate
how accurately these 10 numbers need to be known for the cosmological parameter
accuracies not to be affected appreciably.  We add equal priors to 10 values
$\delta\ln P_i$ ($i=1\ldots 10$) in bins spaced equally in $\log k$ from $k=0.1\hmpcinv$ to
$k=10\hmpcinv$ and marginalize over them, then compute the degradation in the
marginalized accuracies of $\Omega_M$, $\sigma_8$ and $w$.  The bottom panel of
Figure \ref{fig:Pk_prior} shows that a prior on the $P(k)$ values  of a
few percent (2\% for $\sigma_8$, 10\% for $\Omega_M$, 5\% for $w$) would render
the power spectrum essentially ``precisely known'' for a SNAP-type experiment
(1000 sq.\ deg., 100 gal/arcmin$^2$), and enable determination of the three
cosmological parameters without degradations greater than about $10$-$20\%$. The required
accuracies are about two times more stringent for an LSST-type experiment
(15000 sq.\ deg., 30 gal/arcmin$^2$).  Also interesting to note is that
$\sigma_8$ is more degraded than $\Omega_m$ and $w$. This can be easily
understood since, while $\sigma_8$ depends on the lensing power spectra solely
through its dependence on $P(k)$, the parameters $\Omega_{m}$ and $w$ also
enter the lensing geometrical factor that only depends on the cosmic expansion
history. In other words, most information on dark energy parameters comes from
the cosmic geometrical factor -- provided that the redshift distribution of
galaxies is precisely known (see Ma, Hu \& Huterer 2004 and Huterer et al.\ 2004 for a
discussion of the required accuracy of photometric redshifts for weak lensing
tomography).

\section{Systematic biases in $P(k)$}
\label{sec:sys}

We now estimate the bias in cosmological parameters, $\delta p_i$, due to the
systematic bias in $P(k)$. We can easily estimate $\delta p_i$ using the Fisher
matrix formalism.  Let $p_i$ be the true values of cosmological parameters, and
$\bar{C}^{\kappa}_{\alpha}$ and $C^{\kappa}_{\alpha}$ be the true and perturbed
values of the convergence power spectra that result from the bias in
$P(k)$. For clarity we have denoted a pair of redshift bins by a single
subscript $\alpha$.  Then, as long as the perturbations in the convergence
power spectra are small compared to their measurement accuracy, one can derive
the expression to estimate the bias in the cosmological parameters:
\begin{equation}
\delta p_i = 
F_{ij}^{-1} \sum_{\ell}
\left [C_{\alpha}^{\kappa}(\ell)-\bar C_{\alpha}^{\kappa}(\ell)\right ]
\,{\rm Cov}^{-1}\left [\bar{C}^{\kappa}_{\alpha}(\ell), 
  \bar{C}^{\kappa}_{\beta}(\ell)\right ]
\,{\partial \bar C_{\beta}^{\kappa}(\ell) \over \partial p_j},
\label{eq:bias}
\end{equation}
\noindent where $F_{ij}^{-1}$ is the inverse of the Fisher matrix for the
parameters $p_i$, and the summation over $j$, $\alpha$ and $\beta$ is implied.

\subsection{Some examples of realistic biases in $P(k)$}

We first estimate the bias in cosmological parameters due to the systematic
error in $P(k)$ estimated in N-body simulations.  As mentioned earlier, the
only reliable way to make accurate theoretical predictions for WL is to rely on
numerical simulations and perform ray-tracing photons through simulated
large-scale structures (e.g., Jain, Seljak \& White 2000, White \& Hu
2000, Heitmann et al. 2004). 
The accurate model predictions for a given cosmological model
then have to be constructed from a sufficient number
of the simulated lensing map realizations to reduce the statistical variance.  
However, clearly the simulated lensing map will contain uncertainties due to
various numerical limitations.  Vale \& White (2003) derived a useful,
approximate expression for how the finite resolution of the simulation and shot
noise due to finite particle number affect the 3D matter power spectrum:
\begin{equation}
\Delta^2_e(k)=\left(
\Delta^2_m(k)e^{-\sigma_n^2k^2}+\frac{k^3}{2\pi^2\bar{n}}
\right)e^{-\sigma_g^2k^2},
\label{eqn:bpk}
\end{equation}
\noindent where $\Delta_e^2(k)\equiv k^3 P_e(k)/(2\pi^2)$ is an effective 3D
matter power spectrum, $\Delta_m^2(k)$ is the underlying true power spectrum,
$\bar{n}$ is the mean particle number density and $\sigma_n$ and $\sigma_g$ are
characteristic resolution limits of the N-body simulation and the Fourier grid
of the ray-tracing simulation, respectively. Following Vale \& White (2003) we
assume that $\sigma_g$ and $\sigma_n$ can be approximated as
$\sigma_g=0.54L_{\rm box}/N_{\rm grid}$ and $\sigma_n=0.05 \bar{n}^{-1/3}$,
where $N_{\rm grid}^2$ is the ray-tracing grid number 
and $L_{\rm box}$ (Mpc) is
the box size of N-body simulation used. Note that $\bar{n}$ is expressed
in terms of $L_{\rm box}$ and N-body particle number $N$ as 
$\bar{n}=N/L_{\rm box}^3$.
%The equation above shows that
%the numerical error in $P(k)$ appears at $k\simgt {\rm
%min}[\sigma_g^{-1},
%\bar{n}^{1/3}]$.

\begin{figure}
\begin{center}
\includegraphics[height=3.2in, width=4.2in]{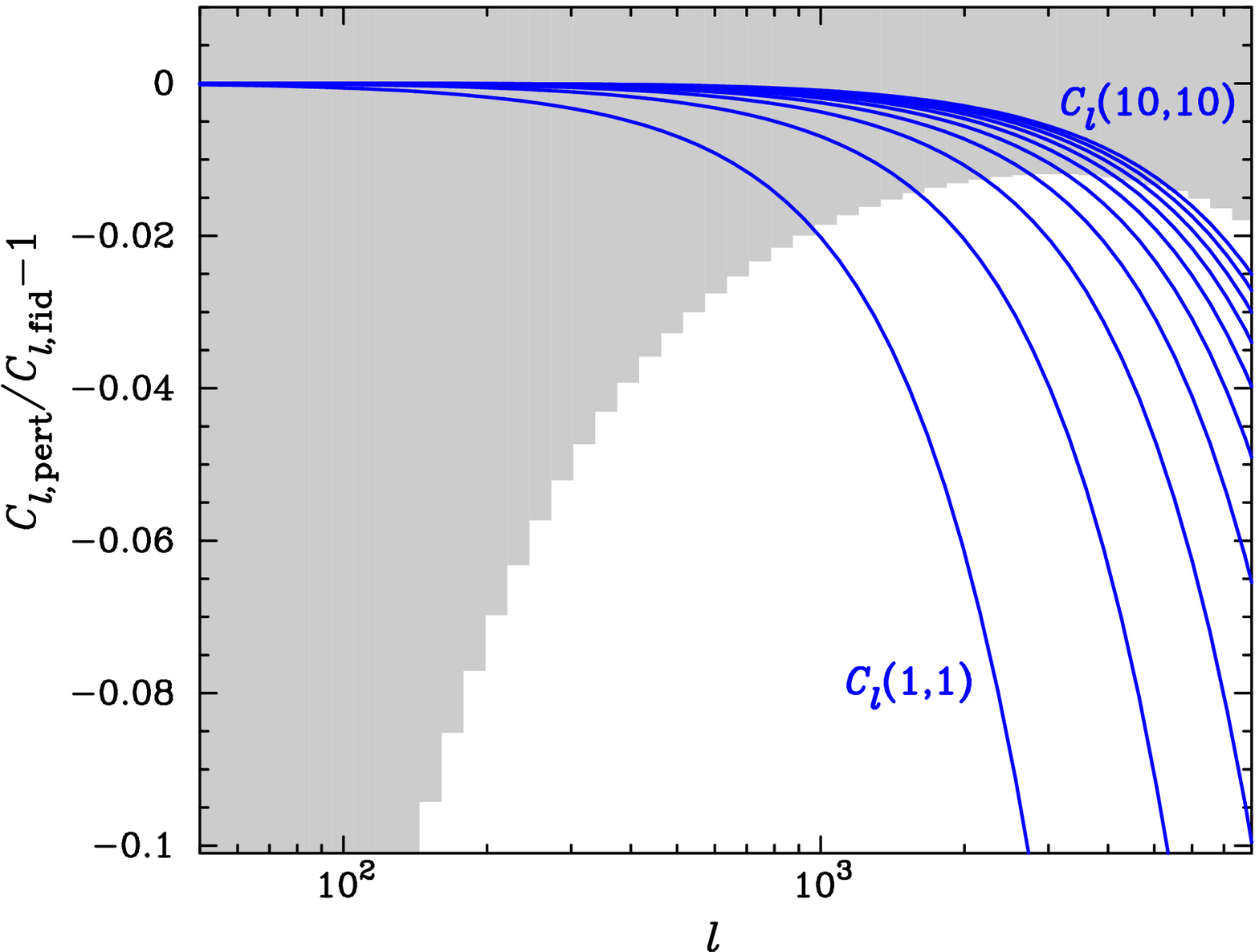}
\includegraphics[height=3.2in, width=4.2in]{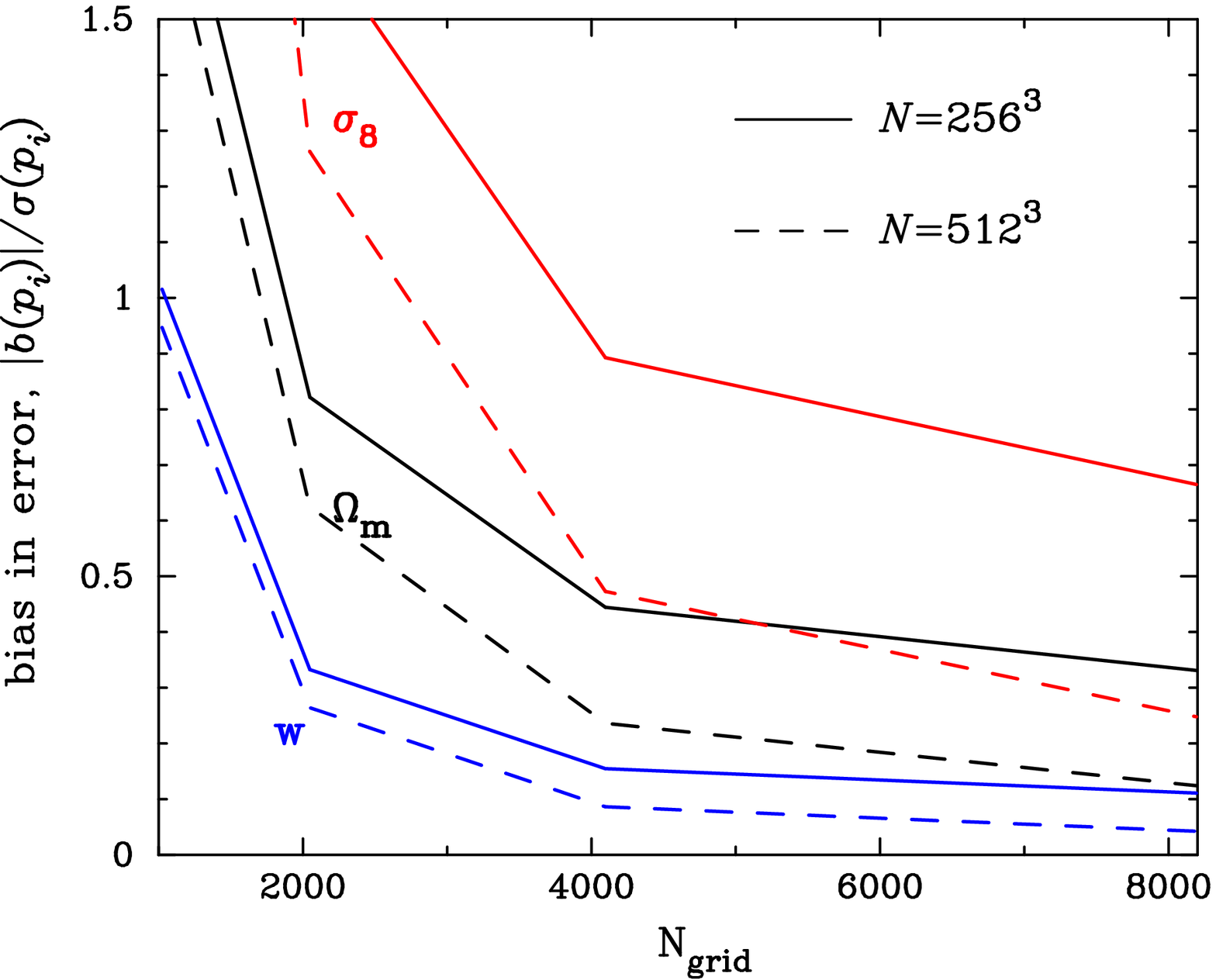}
\end{center}
\caption{The top panel shows fractional differences between the lensing power
spectra with and without the model biases in $P(k)$ due to finite size and
resolution of the simulation, {\it \`{a} la} Eq.~(\ref{eqn:bpk}).  The solid
curves show the auto spectra of 10 redshift bin tomography, where the redshift
bins are equally spaced between $z=0$ and $z=3$ and we assumed the 
simulation parameters of $L=200$ Mpc$/h$, $n=512^3$, and $N_{\rm grid}=4096$.
The shaded region shows the $1\sigma$ statistical error for measuring the 5th
redshift bin spectrum, $C_{55}^{\kappa}$, for the SNAP-type survey
(1000 deg$^2$, 100 gal/arcmin$^{2}$ and
$\langle\gamma_{\rm int}\rangle=0.22$).  Bandwidth of $\Delta \ell/\ell\approx
0.11$ is assumed.  The bottom panel shows how the bias in $P(k)$ propagates
into the bias in cosmological parameters. We plot the bias in the cosmological
parameters relative to their marginalized error vs. the ray-tracing simulation
grid number $N_{\rm grid}$.  The solid and dashed curves show the results for
the two cases of N-body simulation particle number, $N=256^3$ and $N=512^3$,
while the N-body simulation box size is fixed to $200$ Mpc$/h$.}
\label{fig:bias}
\end{figure}
The top panel of Figure \ref{fig:bias} shows the fractional differences between
the convergence power spectra with and without the model numerical error in
$P(k)$. We assumed simulation 
parameters of $L_{\rm box}=200$Mpc$/h$,
$N=512^3$ and $N_{\rm grid}=4096$, and in this case the term including
$\sigma_{n}$ in Eq.~(\ref{eqn:bpk}) which models the finite force resolution in
the N-body simulation provides dominant contribution to the error in $P(k)$
(also see Figure 3 in White \& Vale 2004).  We find that the model error in
$P(k)$ causes the suppression of the lensing power spectrum amplitudes by
$\simlt 10\%$ for $\ell\le 3000$.  Since a given angular mode $\ell$ corresponds
to smaller $k$ as redshift increases, the lensing spectra of higher redshift
bins arise from density perturbations of smaller $k$ and are therefore less
affected by the numerical effects. The shaded region shows the 1$\sigma$
statistical error in the lensing spectrum measurement for the fiducial survey,
where the error arises mainly from the cosmic sample variance at $\ell\simlt
10^3$, while the shot noise due to the intrinsic ellipticities is dominant at
larger $\ell$.  The departures in the lensing spectra of low redshift bins are
larger than the statistical errors even though we only consider the measurements at $\ell\le
3000$.

Thus, it is important to investigate how the numerical inaccuracy in $P(k)$
propagates into the biases in cosmological parameters, and we address this in
the bottom panel of Figure \ref{fig:bias}.  As with the statistical error in
$P(k)$ (see Figure \ref{fig:Pk_prior}), the biases in $P(k)$ lead to a larger bias
in $\sigma_8$ than those in $\Omega_{\rm m}$ and $w$.  It is reassuring to see
that the simulations with $N_{\rm grid}\ge 4096$ and $N=512^3$, which are
feasible with current numerical resources, lead to the bias in $w$ and
$\Omega_{\rm m}$ of less than about 20\% relative to the marginalized error.
Moreover, in the future one can expect to be able to calibrate, and partly correct for, 
the remaining bias in simulations by using a small number of
even higher resolution simulations as a reference. 

The effect of baryons on the non-linear mass clustering is another potentially
harmful systematic effect on $P(k)$ (White 2004, Zhan \& Knox 2004) and its
calibration requires hydrodynamical N-body simulations that include baryon
dissipation.  Baryons, which constitute more than 10\% of the total mass, cool
and condense within a halo, and their effects are most pronounced for halos of
mass $\simlt 10^{13}M_\odot$ (White 2004). The $X$-ray observations indicate
that the hot baryons are likely to have a smoother profile than
dark matter. Zhan \& Knox (2004) estimated the
effect of hot baryons on the lensing power spectrum assuming an isothermal
$\beta$-model to describe the density profile of the baryon within the host
halo.  Their fiducial model leads to modifications in the lensing power
spectrum of a few percent around $\ell\approx 10^3$ for source redshift $z=1$
(see Figure 2 in their paper). This effect is of similar amplitude as that shown in Figure
\ref{fig:bias}, suggesting that the presence of hot baryons
does not necessarily lead to drastic biases in the cosmological
parameters. However, a much more extensive investigation that uses
a suite of hydrodynamical simulations is needed to properly assess the 
effect of baryons on cosmological parameter estimation.

\subsection{Worst-case Systematic Bias in $P(k)$}

The most damaging systematic error in calibrating $P(k)$ will generally be
error that mimics the behavior of cosmological parameters. Somewhat
confusingly, however, the worst-case biased $P(k)$ (call it $Q(k)$) is
typically {\it not} the matter power spectrum for some neighboring values of
cosmological parameters. For, $Q(k)$ is weighted with the geometric factor to
produce the observable convergence power spectrum, and the geometric factor
is known exactly (provided that the redshift distribution of  galaxies is
known). Instead, the worst-case $Q(k)$ will be such that, when weighted by the
geometric factor of the true cosmological model, it produces the convergence power
spectrum which precisely matches some neighboring cosmological model.

While finding the actual worst-case $Q(k)$ is a thorny problem, we can compute
the desired results in the approximation that the perturbation in $P(k)$ is
small. %, based on the Fisher matrix formalism.  
The assumed deviation in the
convergence power spectra is solely due to the bias in the matter power
spectrum
\begin{equation}
C_{\alpha}^{\kappa}(\ell)-\bar C_{\alpha}^{\kappa}(\ell)=
\sum_{a=1}^N \tilde{W}(k_a)\,\delta P(k_a),
\end{equation}
\noindent where we have  discretized the integral as the sum over $N$ lens planes,
or equivalently over $N$ values of the wavenumber $k=\ell/r(z)$, and defined
$\tilde{W}(k)$ appropriately; c.f.\ Eq.~(\ref{eq:pk_l}).  
Then Eq.~(\ref{eq:bias}) can be rewritten as
\begin{equation}
\delta p_i = \sum_{a=1}^N (\delta \ln P(k_a)) \,c_a^{(i)},
\label{eq:delta_w}
\end{equation}
\noindent where we defined
\begin{equation}
c_a^{(i)}\equiv F_{ij}^{-1} \sum_{\ell}
P(k_a)\, \tilde{W}_{\alpha}(k_a)
\,{\rm Cov}^{-1}\left [\bar{C}^{\kappa}_{\alpha}(\ell), 
  \bar{C}^{\kappa}_{\beta }(\ell)\right  ]
\,{\partial \bar C_{\beta}^{\kappa}(\ell) \over \partial p_j},
\label{eq:c}
\end{equation}
\noindent where the summation over $j$, $\alpha$ and $\beta$ is implied.  Let
us assume for a moment that the bias in $\ln P(k)$ is known to be smaller, by
absolute value, than some fixed number $\delta \ln P$. Then
Eq.~(\ref{eq:delta_w}) shows that the worst-case bias in $P(k)$ -- one that
maximizes $\delta p_i$ -- takes the value $\pm (\delta \ln P)$ on each lens
plane, with the sign equal to the sign of $c^{(i)}$ on that plane.  The
resulting worst-case bias in the cosmological parameter is $\delta p_i= (\delta
\ln P) \sum_{a=1}^N |c_a^{(i)}|$.

Turning the argument around, if the bias in a given cosmological
parameter $p_i$ is to be smaller than $\delta p_i$, it is sufficient to
trust the matter power spectrum to an accuracy better than
\begin{equation}
\delta \ln P = {\delta p_i \over \sum_{a=1}^N \,|c_a^{(i)}|}, 
\end{equation}
\begin{figure}
\includegraphics[height=5in, width=4in,angle=-90]{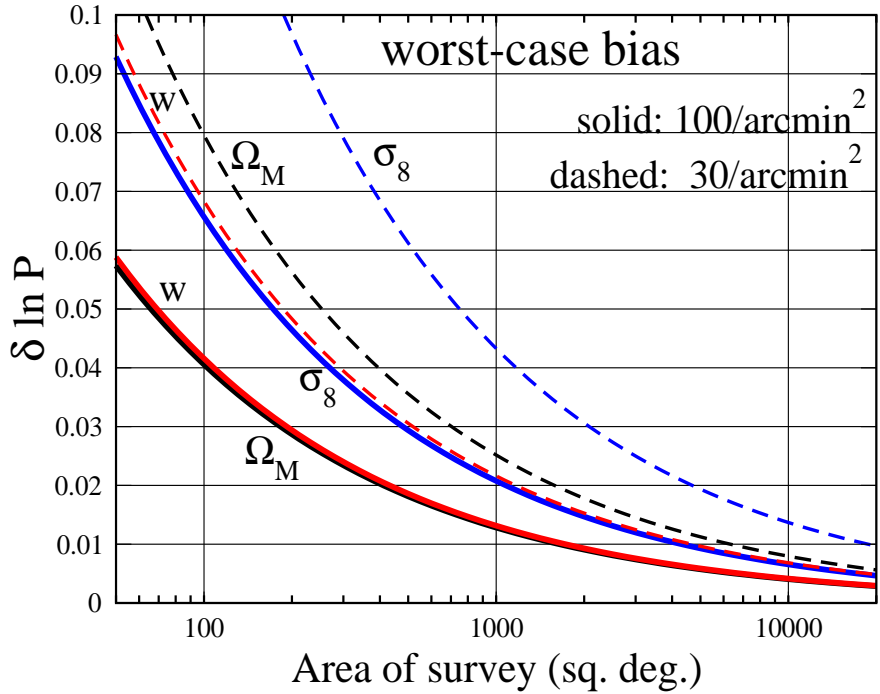}
\caption{The necessary accuracy in $P(k)$ for the worst-case scenario
of bias in $P(k)$, defined so that it produces the maximal bias in the
three cosmological parameters shown. The required accuracy, 
plotted as a function of the sky coverage of the survey, is defined so
that it limits the biases in the cosmological parameters to be no larger than
their 1$\sigma$ fiducial accuracies.
We show the cases of $n_g=100$ and $30$ gal/arcmin$^2$}
\label{fig:worstcase}
\end{figure}
\noindent and this is true irrespective of the conspiring shape of $\delta \ln P(k)$.
Figure \ref{fig:worstcase} shows the required accuracy in $P(k)$, $\delta \ln P$, assuming the
worst-case scenario where the bias equals the 1$\sigma$ accuracy in $p_i$. The
required accuracy is shown as a function of area of the survey for two values
of the density of galaxies (100 and 30 galaxies/arcmin$^2$) and for three
cosmological parameters that are of greatest interest to us ($\Omega_M$,
$\sigma_8$ and $w$).  As before, we find that more ambitious surveys have more
stringent requirements on the systematic control of $P(k)$. We find that, in
order for the worst-case systematics to be subdominant to the 1$\sigma$ error
bars in the three cosmological parameters, a survey like SNAP will need a
1\% calibration in $P(k)$ over the relevant scales, while LSST
needs a better than 0.5\% calibration. We emphasize that these are the required
sensitivities for the {\bf worst-case scenario} of bias in $P(k)$, where the
bias in $P(k)$ conspires to introduce the maximal bias in the cosmological
parameters. Section \ref{sec:stat} shows that a more common error in $P(k)$
that is uncorrelated with cosmological parameters produces biases that are a
factor of a few smaller.

\section{Conclusions}\label{sec:conclusions}

Accurate knowledge of the full nonlinear matter power spectrum is crucial for
the future weak gravitational lensing surveys to achieve their full
potential. In this paper we have explored the required accuracy in $P(k, z)$ so
that it does not bias the cosmological parameter extraction from future
wide-field surveys. This analysis complements other analyses of theoretical and
observational systematics in weak lensing measurements (e.g. Vale et al. 2004,
Huterer et al. 2004).

We find that the power spectrum sensitivity is greatest at two decades in
wavenumber centered around $k=1$ Mpc$/h$ comoving.
%that probe mainly lensing effects on galaxies at $z\sim 1$.  
Calibration of $P(k)$ on those scales needs to be better than a few percent
($\approx 2\%$ for $\sigma_8$ and 5-10\% for $\Omega_M$ and $w$) for the
SNAP-type experiment, and about a factor of two more stringent for the
LSST-type experiment.  We also consider the systematic biases in the power
spectrum in N-body simulations.  While the simulations' accuracy is already
nearly sufficient, the cooling and collapse of baryons, and even the clustering
of massive neutrinos (Abazajian et al.\ 2004) may lead to large errors if they
are not modeled accurately.  There exist a number of options to protect against
such effects (Huterer \& White 2004), but exploring those is beyond the scope
of this work.  The most stringent accuracy requirement on $P(k)$ is obtained
for the worst-case scenario where the bias in $P(k)$ conspires to produce the
convergence power spectrum that is precisely reproduced by cosmological
parameters with biased values. Fortunately, we find that even this worst-case
scenario leads to tolerable biases in the cosmological parameters provided
$P(k)$ is calibrated to 0.5-1\% accuracy over the relevant scales.  Note too
that, in order to achieve $\sim 1$\% accuracy in $P(k)$, current analytic fits
for the {\it linear} power spectrum are not accurate enough either, and one
instead needs to numerically solve the coupled Einstein, fluid and Boltzmann
equations as done e.g. in the CMBFAST code (Seljak \& Zaldarriaga 1996).

As we are entering the era of powerful wide-field WL surveys, currently popular
fitting formulae for the power spectrum will soon become inadequate and one
will need to resort to a full suite of N-body simulations, interpolating on a
grid of cosmological models. First steps in such an approach were recently
undertaken (White \& Vale 2004) although many details remain to be worked
out. We have shown here that the required accuracy is nearly within reach with
current simulations, although the number of models that are necessary to be run
might be very large. With the effort of the weak lensing community and
ever-more powerful computers, the necessary simulation results should be in
hand by the end of the decade, enabling weak lensing to realize its full power
as a probe of dark energy and matter distribution in the universe.

\bigskip

DH is supported by the NSF Astronomy and Astrophysics Postdoctoral Fellowship. 
We thank Kev Abazajian, Gary Bernstein, Salman Habib, 
Bhuvnesh Jain, Lloyd Knox, Eric Linder, 
Chris Vale and Martin White for useful conversations.

\end{document}